\documentclass[twocolumn,preprint2]{aastex631}

\usepackage{epsfig}
\usepackage{diagbox}
\usepackage{multirow}

\begin{document}

\title{Evolution Models of CO WD - AGB Star Merger Remnants}

\author[0000-0002-2452-551X]{Chengyuan Wu}
\affiliation{Yunnan Observatories, Chinese Academy of Sciences, Kunming 650216, China}
\affiliation{International Centre of Supernovae, Yunnan Key Laboratory, Kunming, 650216, China}
\email{wuchengyuan@ynao.ac.cn}

\author{Heran Xiong}
\affiliation{Research School of Astronomy and Astrophysics, The Australian National University, Canberra, ACT 2611, Australia}

\author{Shi Jia}
\affiliation{Yunnan Observatories, Chinese Academy of Sciences, Kunming 650216, China}
\affiliation{International Centre of Supernovae, Yunnan Key Laboratory, Kunming, 650216, China}

\author[0009-0006-8370-3108]{Zhengyang Zhang}
\affiliation{Yunnan Observatories, Chinese Academy of Sciences, Kunming 650216, China}
\affiliation{International Centre of Supernovae, Yunnan Key Laboratory, Kunming, 650216, China}
\affiliation{University of Chinese Academy of Sciences, Beijing 100049, China}

\author{Bo Wang}
\affiliation{Yunnan Observatories, Chinese Academy of Sciences, Kunming 650216, China}
\affiliation{International Centre of Supernovae, Yunnan Key Laboratory, Kunming, 650216, China}
\email{wangbo@ynao.ac.cn}

\begin{abstract}

Common envelope evolution is a critical but still poorly understood phase in binary evolution. It plays a key role in forming close binaries such as hot subdwarfs, double white dwarfs, X-ray binaries, and double neutron stars. However, its outcomes remain highly uncertain. Depending on the efficiency of envelope ejection, a system may either survive as a close binary or undergo a complete merger. In this work, we investigate the post merger evolution of systems where a CO WD mergers with the core of an AGB star. A grid of merger remnant models with various core and envelope masses is constructed. At the onset of evolution, the CO core contracts and undergoes off-center carbon ignition, producing an inwardly propagating carbon flame. For remnants with relatively low mass of CO core, the flame phase is followed by core contraction and subsequent H-shell burning. For more massive CO cores, the carbon flame reaches the center and is soon followed by off-center neon burning, which is expected to eventually lead to core-collapse supernovae. The merger remnants occupy nearly the same region on HR diagram as ordinary AGB or super-AGB stars, exhibiting similar surface properties. Although their surface abundance may differ slightly from those of normal AGB stars depending on the initial core and envelope masses, these differences are strongly reduced once mass-loss is taken into account. We suggest that some giant-like stars, including candidates for Thorne-${\rm {\dot Z}{ytkow}}$ objects (e.g., HV 2112), might alternatively be explained as AGB-WD merger remnants.

\end{abstract}


\section{Introduction}

In recent years, an increasing number of close binary systems have been discovered thanks to large-scale sky surveys, such as Zwicky Transient Facility (e.g., \citealt{2019PASP..131a8003M}) high-cadence Galactic plane survey (e.g., \citealt{2021MNRAS.505.1254K}), OmegaWhite (e.g., \citealt{2015MNRAS.454..507M}), Tsinghua University–Ma Huateng Telescopes for Survey (e.g., \citealt{2020PASP..132l5001Z}; \citealt{2022MNRAS.509.2362L}), and the Wide Field Survey Telescope (e.g., \citealt{2023SCPMA..6609512W}; \citealt{2025ApJS..278...29L}). One of the most widely accepted mechanisms for the formation of such close binaries is the common envelope evolution (CEE). The CE phase is a short stage in binary evolution, during which the two stars orbit within a shared gaseous envelope. Due to tidal drag, the binary system transfers orbital energy into the envelope, potentially resulting in the envelope ejection. If envelope ejection is successful, the system can survive as acompact binary, producing objects such as cataclysmic variables, white dwarf (WD) binaries, double WDs, and other compact binaries hosting degenerate stars (e.g., \citealt{2020RAA....20..161H}; \citealt{2024PrPNP.13404083C}). CEE is also implicated in several major astrophysical phenomena, including Type Ia supernovae (e.g., \citealt{1984ApJ...277..355W}; \citealt{1984ApJS...54..335I}; \citealt{1985ApJ...297..531N}), nova eruptions (e.g., \citealt{1990ApJ...356..250L}), X-ray binaries (e.g., \citealt{2000ARA&A..38..113T}), double WD mergers (e.g., \citealt{1985A&A...150L..21S}; \citealt{2010Natur.463...61P}; \citealt{2013MNRAS.429.1425R}), gravitational wave sources (e.g., \citealt{2002ApJ...572..407B}), and possibly some planetary nebulae (e.g., \citealt{2016ApJ...832..125H}; \citealt{2016MNRAS.455.3511S}). Understanding the outcomes of CEE is therefore essential for constraining the progenitors of Type Ia supernovae, compact object binaries, and stellar-mass gravitational wave sources.

It is believed that the formation of a CE is associated with dynamically unstable mass transfer in binaries (e.g., \citealt{1976IAUS...73...75P}). When the mass ratio exceeds a critical threshold, mass loss causes the donor star to expand faster, disrupting the hydrostatic equilibrium and driving the mass transfer process on a dynamical timescale. In this case, the companion is rapidly engulfed and initiating CEE (e.g., \citealt{2013A&ARv..21...59I}). The CE phase is highly complex, making it one of the most uncertain process in astrophysics. Once the CE forms, the companion spirals inward, depositing orbital energy into the envelope, leading to its expansion and ejection of the envelope. However, several key issues remain debated, including how much orbital energy is required to eject the envelope, and whether additional energy sources, such as hydrogen or helium recombination (e.g., \citealt{1994MNRAS.270..121H}; \citealt{2015MNRAS.447.2181I}; \citealt{2022MNRAS.516.4669L}), dust formation (e.g., \citealt{2020MNRAS.497.3166I}), accretion onto the companion during CEE (e.g., \citealt{2018MNRAS.480.1898C}), or jets launched by the companion (e.g., \citealt{2019MNRAS.488.5615S}), play significant roles in ejecting the envelope.


Because of its short timescale, directly observations of the CE phase are extremely challenging. Nonetheless, some evidence suggests there exists systems that have experienced CE ejection. For example, circumbinary disks or diffuse circumstellar material observed around short-period hot subdwarf binaries may be remnants of CE ejection (e.g., \citealt{2011A&A...531A.158M}; \citealt{2022MNRAS.515.3370L}; \citealt{2025MNRAS.537.1950L}). If the envelope is not fully ejected, the binary will eventually merge within the CEE. In this case, the remnant may retain an extended envelope and resemble a giant star, but with structural and chemical properties potentially different from those of a normal giant (e.g., \citealt{2021MNRAS.508.1618R}). Therefore, exploring the evolution and evolution of CE merger remnants is crucial for assessing this alternative evolutionary channel.

Multi-dimensional hydrodynamic simulations have provided valuable insights into CEE (e.g., \citealt{1998ApJ...500..909S};  \citealt{2016MNRAS.455.3511S}; \citealt{2020A&A...644A..60S}). However, these simulations remain extremely challenging due to the large discrepancy in spatial resolution between the extended envelopes and the compact cores, and is difficult in modeling thermal readjustment and complete envelope ejection process. Most 3D simulations only capture the early spiral-in phase (e.g., over tens of orbital periods) and cannot follow the post-CE evolution. To circumvent these computational limitations, simplified approaches are often adopted, for example one may use energy-balance equations that compare orbital energy with envelope binding energy to estimate post-CE orbital parameters (e.g., \citealt{1984ApJS...54..335I}), and then compare the binary population synthesis results with the observed binary orbital distributions to constrain the CE ejection efficiency. Previous works also used 1D stellar models to study the impact of the release of orbital energy or gravitational potential of the binary star on the primary star, avoiding modeling the spiral-in process directly (e.g., \citealt{2015MNRAS.447.2181I}; \citealt{2017MNRAS.470.1788C}).

Recently, growing attention has been paid to the merger channel itself. 1D models of CE merger remnants have suggested that such systems can be distinguished from ordinary stars through their asteroseismological signals. For example, \cite{2021MNRAS.508.1618R} studied RGB + MS mergers and found significant difference in gravity-mode period spacing, implying that asteroseismology can be a useful tool in identifying merger remnants. They further showed that the He WD + MS merger remnants yield asteroseismology features distinct from normal RGB stars (e.g., \citealt{2024OJAp....7E..81R}). Other studies demonstrated that He WD + RGB mergers can produce Li-rich giants or carbon stars \cite{2020ApJ...889...33Z}, while He WD + MS mergers may lead to the formation of blue large-amplitude pulsators \cite{2023ApJ...959...24Z}. More recently, \cite{2025ApJ...987..212P} modeled Hertzsprung gap stars + MS mergers, showing that their remnants evolve as blue supergiants phase with reduced He core masses and lower carbon abundance compared to stripped stars.

Despite these advances, the role of recombination energy in CEE remains debated. For example, \cite{2020A&A...644A..60S} found that although the envelope of an AGB star is less tightly bound than that of a red giant, envelope ejection stalls below 20\% of its mass if ionization energy release is not included. This implies that the companion may continue spiraling inward and potentially merge with the AGB core during CEE. Such merges could significantly alter the structure and evolution of the star, producing remnants with properties distinct from ordinary AGB stars. In this work, we aim to construct a simplified 1D model of merger remnants formed by the coalescence of a CO WD with the core of an AGB star. We then follow their subsequent evolution to assess whether they exhibit observable characteristics that distinguish them from normal AGB stars.

This paper is organized as follows: In Sect.\,2, we describe our methods for constructing the merger models, and present the results of their post merger evolution in Sect.\,3. In Sect.\,4, we discuss the implications and limitations of our models. Finally, the summary is provided in Sect.\,5.


\section{Constructing 1D merger models}

To construct 1D model of merger remnants and investigate their post-merger evolution, the ideal approach would be to perform 3D hydrodynamic simulations of the merger process and then map the resulting structure into 1D profiles suitable for stellar evolution codes. This strategy has been widely adopted in studies of double WD mergers. For instance, \cite{2016MNRAS.463.3461S} and \cite{2021ApJ...906...53S} followed this approach to explore the evolution of double CO WD merger remnants by using the 3D merger structures from \cite{2014MNRAS.438...14D}. More recently, \cite{2023ApJ...944L..54W} and \cite{2023MNRAS.525.6295W} extended the method to investigated the evolution of ONe + CO WD and double ONe WD merger remnants.

However, no 3D simulations are currently available for mergers between an AGB star and a CO WD, so there is no ready-made 1D profile to adopt. We therefore make the following simplifying assumptions: (1) Since both the core of AGB star and the CO WD are degenerate, we assume that their merged structure resembles that of a double CO WD merger remnant. Thus, the remnant is assigned a CO core whose mass equals to the sum of the original AGB core and the CO WD, and a temperature profile similar to that of double CO WD merger remnants. This assumption is motivated by the results of \cite{2014MNRAS.438...14D}, who showed that a double WD merger will form a remnant composed of a cool core, a hot envelope, a Keplerian disk, and a tidal tail from the inside out. The density of the hot envelope in a double WD merger remnant ranges from ${10}^{6}$ - ${10}^{5}\,{\rm g}/{\rm cm}^{3}$ from the inside out, while the density of the disk ranges from ${10}^{5}$ - ${10}^{2}\,{\rm g}/{\rm cm}^{3}$. In contrast, the H-rich envelope of an AGB star has a typical density $\sim{10}^{-5}\,{\rm g}/{\rm cm}^{3}$, which is many orders of magnitude lower than either component of the double WD merger remnant. Therefore, we estimate that the presence of the AGB envelope does not significantly affect the structure of the merged core. (2) We further assume that during the spiral-in phase, the CO WD loses orbital energy only to unbind a portion of the AGB envelope, without altering the elemental abundance distribution of the envelope. In other words, we neglect the possible mixing process between the envelope and the WD material. We acknowledge that the merger process may reality affect the structure of the AGB envelope, potentially causing, for example, additional mixing or mass ejection, but such effects are beyond the scope of the present work.

Under the above two assumptions, the merger remnant consists of a massive CO core with a ``double-WD-merger-like'' thermal structure, surrounded by a He-rich inter shell and a H-rich envelope that retains the profile of the original AGB star. In the present work we consider only merger processes occurring during the late AGB phase, for which the He-rich inter shell is extremely thin ($\sim0.01{M}_{\odot}$). In this case, the merger process is unlikely to significantly modify the structure of the He-rich shell. However, for mergers involving a CO WD and an early-AGB star, the He-rich inter shell is more massive than in the late AGB case, the merger process may induce additional mixing between the He-rich shell and the H-rich envelope. Such situation is not considered in the present work.

Based on the basic assumptions describe above, we use stellar evolution code {\tt\string MESA} (version 12778; e.g., \citealp{2011ApJS..192....3P, 2013ApJS..208....4P, 2015ApJS..220...15P, 2018ApJS..234...34P, 2019ApJS..243...10P}) to construct the merger remnant. To reduce the computational cost, we restrict the core masses of AGB star to the range $0.5{M}_{\odot}$-$1.0{M}_{\odot}$, with steps of $0.1{M}_{\odot}$. These AGB stars correspond to progenitor main-sequence (MS) stars of $3.0$-$8.0{M}_{\odot}$, such that ${M}_{\rm {core}}=0.5{M}_{\odot}$ originates from a $3.0{M}_{\odot}$ MS star, ${M}_{\rm {core}}=0.6{M}_{\odot}$ from a $4.0{M}_{\odot}$ MS star, and forth up to ${M}_{\rm {core}}=1.0{M}_{\odot}$ from an $8.0{M}_{\odot}$ MS star. For each case, the CO WD companion is assumed to have the same mass as the core of AGB star. This choice is motivated by binary evolution pathways: CO WD + AGB star binaries naturally form through the stable RLOF + CE channel (e.g., \citealt{2020RAA....20..161H}), and binary population synthesis calculations indicate that, at the onset of the CE phase, the CO WD mass is often comparable to the core mass of the AGB companion (e.g., \citealt{2023A&A...669A..82L}). Therefore, adopting ${M}_{\rm WD}={M}_{\rm core}$ represents the most common configuration expected for CO WD + AGB core mergers. In principle, a more complete model grid should include systems with different mass ratios between the WD and AGB core. However, under the assumption that the core structure of the merger remnant resembles that of double WD mergers, we expect that whether the AGB core is more or less massive than the accreted CO WD does not significantly modify the resulting merger structure. Therefore, for computational efficiency, we assumed the CO WD and the core of the AGB star to have the same mass in the present work, and defer a detailed exploration of varying mass ratio to future studies.

Since the fraction of the AGB envelope lost during the merger process is highly uncertain, we simply consider four limiting cases: (1) no envelope mass-loss during merger process; (2) $50\%$ of envelope mass-loss; (3) $90\%$ of envelope mass-loss; (4) extremely case with $99\%$ of envelope mass-loss. This setup yield a grid of $24$ merger models with different CO core and envelope masses.

The construction procedure involves three main steps, which we illustrate with the case of an $8.0{M}_{\odot}$ AGB star merging with an $1.0{M}_{\odot}$ CO WD. (1) We first evolve an $8.0{M}_{\odot}$ MS star of metallicity ${z}=0.02$ to the AGB phase. We consider ``${\tt {approx21.net}}$'' in ``${\tt {MESA\_default}}$'' as the nuclear reaction network, which includes $21$ isotopes from $^{\rm 1}{\rm H}$ to $^{\rm 56}{\rm Fe}$. Wind mass-loss process is neglected in this illustrative case. The calculation is stopped when the core mass of AGB star reaches $1.0{M}_{\odot}$, at which point the star has an $1.0{M}_{\odot}$ CO core, an extremely thin He-rich inter shell (typically around $0.01{M}_{\odot}$) and a $\sim7.0{M}_{\odot}$ H-rich envelope. The corresponding elemental abundance distribution and entropy profile of the AGB star is shown in panel (a) of Fig.\,1. (2) Following \cite{2021ApJ...906...53S}, we construct the thermal structure of a $2.0{M}_{\odot}$ CO core representative of a double $1.0{M}_{\odot}$ CO WD merger remnant. We evolve a $2.0{M}_{\odot}$ pre-He-MS star (a phase analogous to the pre-MS phase) with metallicity ${z}=0.02$ until its central density reaches ${\rho}_{\rm c}=5.0{\rm g}/{\rm {cm}^{3}}$, relax its elemental abundance distribution to match those of the AGB core, and then adjust its entropy profile by injecting energy in different mass zones until it resembles that of a double WD merger remnant\footnote{The entropy of an AGB core of a given mass is indeed higher than that of a CO WD owing to ongoing shell burning. In our $8.0{M}_{\odot}$ AGB model with an $1.0{M}_{\odot}$ CO core, the core radius is about ${\rm log}({R}_{\rm core}/{R}_{\odot}) = -2.0$, larger than the corresponding CO WD radius, ${\rm log}({R}_{\rm WD}/{R}_{\odot}) = -2.1$ (Eggleton’s zero-temperature mass-radius relation; e.g., \citealt{1988ApJ...332..193V}). This suggests that the AGB core is more likely to be disrupted during the merger. \cite{2013MNRAS.430.2113Z}, who studied mergers between a He WD and a RGB core, found that although the entropy profiles of merger remnants can differ depending on which component is disrupted, the overall structure of the remnant is primarily determined by the mass of the undisrupted core. In the case of equal-mass core mergers, they showed that whether the donor or the accretor core is disrupted does not significantly alter the final merger structure. By analogy, we therefore expect that mergers between an AGB core and a CO WD produce core structures broadly similar to those of double WD mergers.}. The resulting profiles are shown in panel (b) of Fig.\,1. (3) We combine the $2.0{M}_{\odot}$ CO core profile with the $7.0{M}_{\odot}$ envelope profile of the AGB star to create a $9.0{M}_{\odot}$ composite structure. Specifically, the inner $2.0{M}_{\odot}$ part adopts the abundance and entropy profiles of the double WD merger remnant, while the outer $7.0{M}_{\odot}$ part retains the envelope structure of the AGB model. We then evolve a $9.0{M}_{\odot}$ pre-MS star until ${\rho}_{\rm c}=3.0{\rm g}/{\rm {cm}^{3}}$, and relax its abundance distribution to the target profile. Afterwards, we continue evolve the remnant until ${\rho}_{\rm c}=5.0{\rm g}/{\rm {{cm}^{3}}}$ and finally relax its entropy profile. The resulting merger remnant structure is shown in panel (c) of Fig.\,1.

By repeating this procedure for AGB progenitors of $3.0{M}_{\odot}$-$7.0{M}_{\odot}$ (core masses of $0.5{M}_{\odot}$-$0.9{M}_{\odot}$), and applying envelope mass-loss prescriptions, we obtain the full set of $24$ initial merger models. In particular, the reduced envelope cases are constructed by evolving the progenitor AGB stars to the desired core masses and then artificially stripping their envelopes with an enhanced wind mass-loss rate (${10}^{-3}\,{M}_{\odot}\,{\rm {yr}^{-1}}$, which is sufficient given the typical timescale of the CE phase) until the target envelope masses are reached. The details of these models are summarized in the first four columns in Table.\,1.

\begin{figure*}
\begin{center}
\epsfig{file=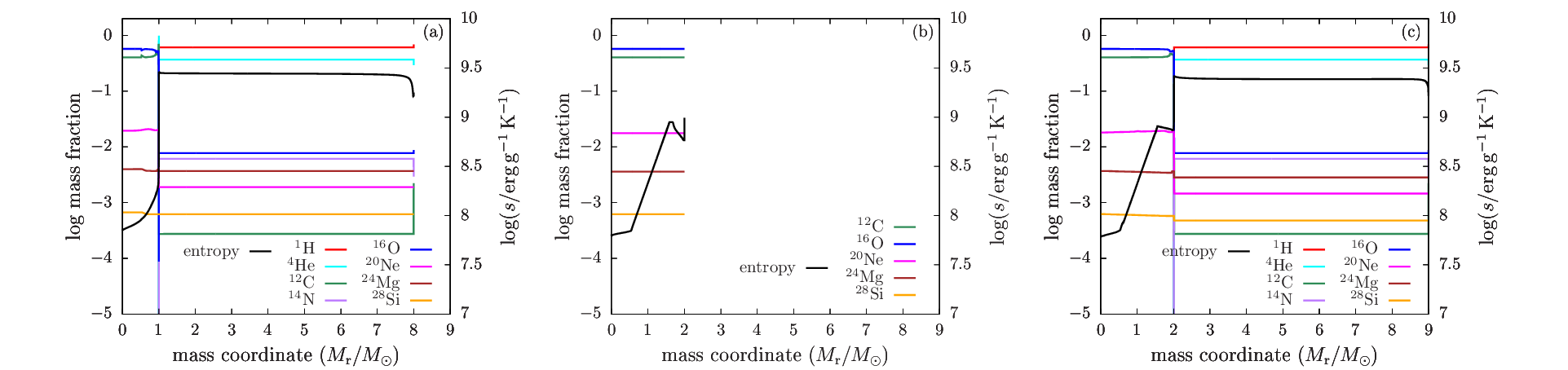,angle=0,width=18.2cm}
 \caption{Structure profiles for constructing the merger remnant of an $8.0{M}_{\odot}$ AGB star and a $1.0{M}_{\odot}$ CO WD. Panel (a): elemental abundance distribution and entropy profile of an $8.0{M}_{\odot}$ AGB star when CO core mass reaches $1.0{M}_{\odot}$. Panel (b): elemental abundance distribution and entropy profile of a double $1.0{M}_{\odot}$ CO WD merger remnant. Panel (c): elemental abundance distribution and entropy profile of the merger remnant formed from the combination of an $8.0{M}_{\odot}$ AGB and a $1.0{M}_{\odot}$ CO WD.}
  \end{center}
    \label{fig: 1}
\end{figure*}

\begin{table*}[htbp]
\centering
\begin{tabular}{|c|c|c|c|c|c|c|}
\hline
\multicolumn{7}{|c|}{AGB + CO WD} \\
\hline
 & & & & & & \\
{Model} & ${M}_{\rm remnant}$ & ${M}_{\rm {AGB,\,progenitor}}$ & ${M}_{\rm {core}}={M}_{\rm {WD}}$ & ${M}_{\rm {env}}$ & Stopping condition & ${\rm t}/{\rm yr}$ \\
 & & & & & & \\
\hline
$3.0+0.5$ & $3.5$ & $3.0$ & $0.5$ & $2.5$ & Shell burning & $8.5\times{10}^{4}$ \\
$1.75+0.5$ & $2.25$ & $3.0$ & $0.5$ & $1.25$ & Shell burning & $4.57\times{10}^{4}$ \\
$0.75+0.5$ & $1.25$ & $3.0$ & $0.5$ & $0.25$ & Shell burning & $6.0\times{10}^{4}$ \\
$0.525+0.5$ & $1.025$ & $3.0$ & $0.5$ & $0.025$ & Shell burning & $6.8\times{10}^{4}$ \\
\hline
$4.0+0.6$ & $4.6$ & $4.0$ & $0.6$ & $3.4$ & Shell burning & $3.47\times{10}^{4}$ \\
$2.3+0.6$ & $2.9$ & $4.0$ & $0.6$ & $1.7$ & Shell burning & $3.1\times{10}^{4}$ \\
$0.94+0.6$ & $1.54$ & $4.0$ & $0.6$ & $0.34$ & Shell burning & $5.38\times{10}^{4}$ \\
$0.634+0.6$ & $1.234$ & $4.0$ & $0.6$ & $0.034$ & Shell burning & $5.64\times{10}^{4}$ \\
\hline
$5.0+0.7$ & $5.7$ & $5.0$ & $0.7$ & $4.3$ & Shell burning & $2.66\times{10}^{4}$ \\
$2.85+0.7$ & $3.55$ & $5.0$ & $0.7$ & $2.15$ & Shell burning & $2.07\times{10}^{4}$ \\
$1.13+0.7$ & $1.83$ & $5.0$ & $0.7$ & $0.43$ & Shell burning & $2.68\times{10}^{4}$ \\
$0.743+0.7$ & $1.443$ & $5.0$ & $0.7$ & $0.043$ & Shell burning & $2.85\times{10}^{4}$ \\
\hline
$6.0+0.8$ & $6.8$ & $6.0$ & $0.8$ & $5.2$ & Off-center Ne burning & $1.14\times{10}^{4}$ \\
$3.4+0.8$ & $4.2$ & $6.0$ & $0.8$ & $2.6$ & Off-center Ne burning & $1.26\times{10}^{4}$ \\
$1.32+0.8$ & $2.12$ & $6.0$ & $0.8$ & $0.52$ & Shell burning & $1.25\times{10}^{4}$ \\
$0.852+0.8$ & $1.652$ & $6.0$ & $0.8$ & $0.052$ & Shell burning & $1.21\times{10}^{4}$ \\
\hline
$7.0+0.9$ & $7.9$ & $7.0$ & $0.9$ & $6.1$ & Off-center Ne burning & $1.08\times{10}^{4}$ \\
$3.95+0.9$ & $4.85$ & $7.0$ & $0.9$ & $3.05$ & Off-center Ne burning & $1.13\times{10}^{4}$ \\
$1.51+0.9$ & $2.41$ & $7.0$ & $0.9$ & $0.61$ & Off-center Ne burning & $1.05\times{10}^{4}$ \\
$0.961+0.9$ & $1.861$ & $7.0$ & $0.9$ & $0.061$ & Off-center Ne burning & $1.08\times{10}^{4}$ \\
\hline
$8.0+1.0$ & $9.0$ & $8.0$ & $1.0$ & $7.0$ & Off-center Ne burning & $8.33\times{10}^{3}$ \\
$4.5+1.0$ & $5.5$ & $8.0$ & $1.0$ & $3.5$ & Off-center Ne burning & $9.2\times{10}^{3}$ \\
$1.7+1.0$ & $2.7$ & $8.0$ & $1.0$ & $0.7$ & Off-center Ne burning & $8.77\times{10}^{3}$ \\
$1.07+1.0$ & $2.07$ & $8.0$ & $1.0$ & $0.07$ & C flame almost reaches center & $7.38\times{10}^{3}$ \\
\hline
\end{tabular}
\caption{\label{tab:1} Information of the $24$ fiducial merger models. Columns $1$-$7$ represent, (1) Model: masses of the progenitor AGB star and the CO WD; (2) ${M}_{\rm {remnant}}$: total mass of the merger remnant (${M}_{\odot}$); (3) ${M}_{\rm {AGB,\,progenitor}}$: Initial mass of the AGB progenitor; (4) ${M}_{\rm {core}}={M}_{\rm {WD}}$: CO core mass of the AGB star at merger, equal to the mass of the CO WD; (5) ${M}_{\rm {env}}$: Envelope mass of the AGB star at merger; (6) Stopping condition: evolutionary stage at which the calculation was terminated; (7) ${\rm t}/{\rm yr}$: evolutionary time of the merger remnant.} 
\end{table*}

\section{Evolution of merger remnant}

After obtaining the initial merger models, we restored the nuclear burning and mixing processes to evolve the merger remnants over time. The nuclear reaction network used in this work is ``approx21.net'', which include $21$ isotopes from $^{\rm 1}{\rm H}$ to $^{\rm 56}{\rm Fe}$ and their corresponding nuclear chains, covering most of the key nuclear reactions during stellar evolution. Given that the merger remnants are composed of metal-rich material, we used the OPAL type $2$ opacity table to calculate their evolution (e.g., \citealt{1996ApJ...464..943I}). Since the merger process increases the core mass of the AGB stars, off-center carbon burning may be triggered in the CO core, similar to what occurs in double CO WD merger remnants (e.g., \citealt{2021ApJ...906...53S}). To better resolve the propagation of the off-center flame, we adjusted the spatial resolution by setting ``${\tt {dlog\_burn\_c\_dlogP\_extra=0.1}}$'' to refine the mesh grid around the carbon flame. For remnants with a massive CO core, the off-center carbon flame propagates inward, eventually reaching the center and triggering the off-center Ne burning. To save the computation time, we stopped the simulations when the off-center Ne burning occurs. For less massive merger remnants, the core mass is insufficient to trigger the off-center Ne burning. Thus, we halted the calculations when the carbon flame reached the center and initiated the shell carbon burning. The stopping conditions and evolutionary times for each model are summarized in the last two columns of Table.\,1. Some physical processes, such as wind mass-loss process and stellar rotation, are not included in our fiducial models. We will discuss the effects of these uncertainties in sect.\,4.

\begin{figure*}
\begin{center}
\epsfig{file=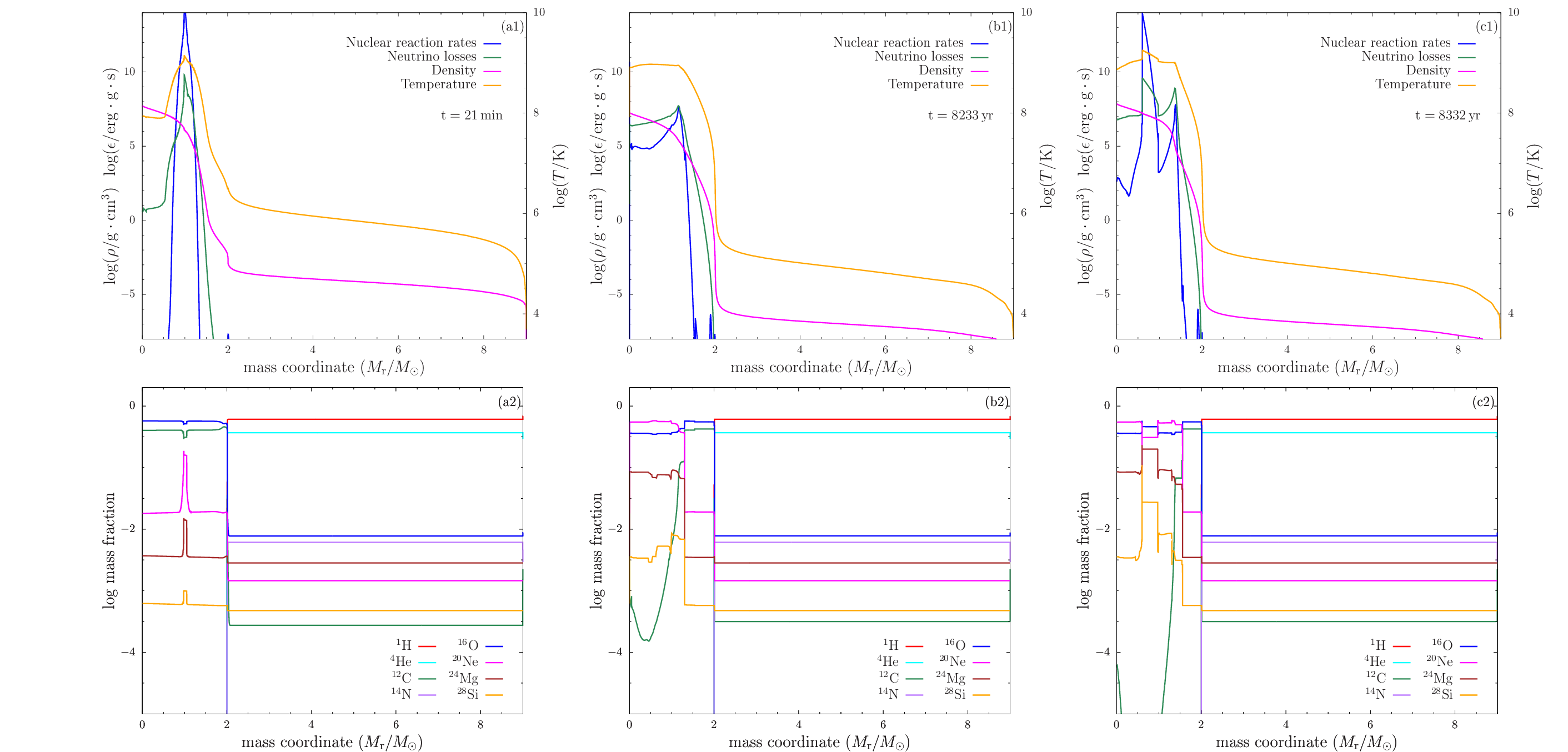,angle=0,width=18.2cm}
 \caption{Structure profiles of the $8.0{M}_{\odot}$ AGB star + $1.0{M}_{\odot}$ CO WD merger remnant at different evolutionary stages. Panel (a1) and (a2): temperature, density, nuclear energy generation rate, and neutrino cooling rate (a1), and elemental abundance distribution (a2) at the onset of off-center carbon ignition; Panel (b1) and (b2): same as above, when the inwardly propagating carbon flame reaches the center; Panel (c1) and (c2): same as above, at the onset of off-center neon ignition.}
  \end{center}
    \label{fig: 2}
\end{figure*}

\begin{figure*}
\begin{center}
\epsfig{file=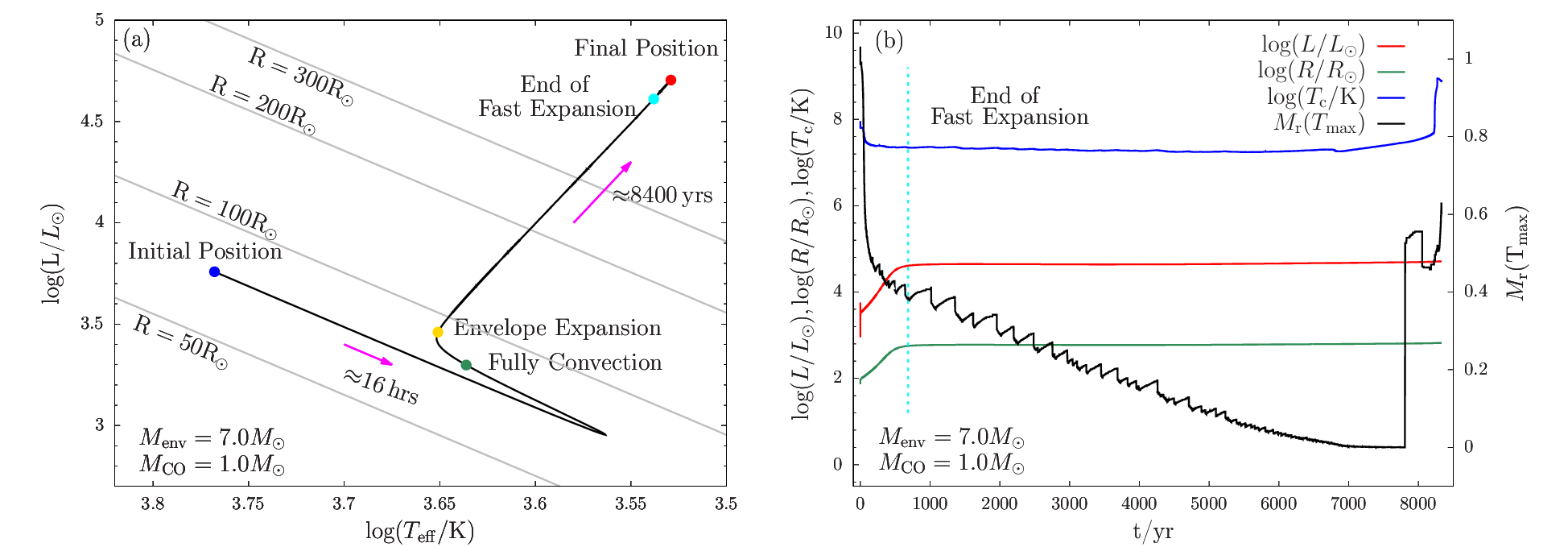,angle=0,width=16.2cm}
 \caption{Evolution of the $8.0{M}_{\odot}$ AGB star + $1.0{M}_{\odot}$ CO WD merger remnant. Panel (a): HR diagram, where blue, golden, sea-green, cyan and red filled cycles mark different evolutionary stages. Magenta arrows indicate the evolutionary directions, with the corresponding timescale labeled. Gray diagonal lines denote constant stellar radii. Panel (b): time evolution of luminosity (red), radius (sea-green), central temperature (blue), and the mass coordinate of the maximum temperature (black). Cyan vertical dotted line marks the end of rapid envelope expansion phase, corresponding to the cyan filled cycle in panel (a).}
  \end{center}
    \label{fig: 3}
\end{figure*}

We present the evolution of the most massive merger remnant (i.e., $8.0{M}_{\odot}$ AGB star + $1.0{M}_{\odot}$ CO WD) as an example. The structure and abundance profiles at various evolutionary stages are shown in Fig.\,2, while the evolution of the remnant on HR diagram and the evolution of other key information are present in Fig.\,3. At the onset of evolution, the core begins to contract, triggering off-center carbon burning (see also \citealt{2016MNRAS.463.3461S}; \citealt{2021ApJ...906...53S}. Given the high initial temperature of the core, carbon burning is ignited within approximately $21$ minutes. The maximum temperature occurs at a mass coordinate of about ${M}_{\rm r}\approx0.985{M}_{\odot}$. During the core contract phase, H shell burning is temporarily quenched, which leads to a decrease in the temperature of the remnant. After about $16$ hours of off-center carbon burning, the released nuclear energy is transported outward, causing both the effective temperature and luminosity of the remnant to rise. Once the H-rich envelope becomes fully convective, the remnant enters the phase of rapid expansion. The off-center carbon flame converts $^{\rm 12}{\rm C}$ into $^{\rm 20}{\rm Ne}$ and $^{\rm 24}{\rm Mg}$, and the flame starts to propagate inward. During the carbon burning phase, the remnant ascends along the AGB branch on the HR diagram. As the carbon flame moves inward, the position of the maximum temperature shifts accordingly. It takes approximately $8400$ years for the flame to reach the center. When the flame reaches the center, carbon burning ceases, leading to the contraction of the ONe core. At this point, the position of the maximum temperature moves outward, around ${M}_{\rm r}\approx0.6{M}_{\odot}$. Off-center Ne burning then begins, producing an inwardly propagating Ne flame. This transition is shown in the black line in panel (b) of Fig.\,3. We stop the calculation when the off-center Ne burning starts. We infer that the Ne flame will reach the center in about $20$ years (e.g., \citealt{2023MNRAS.525.6295W}), at which point the remnant will explode as a core collapse supernova.

\begin{figure*}
\begin{center}
\epsfig{file=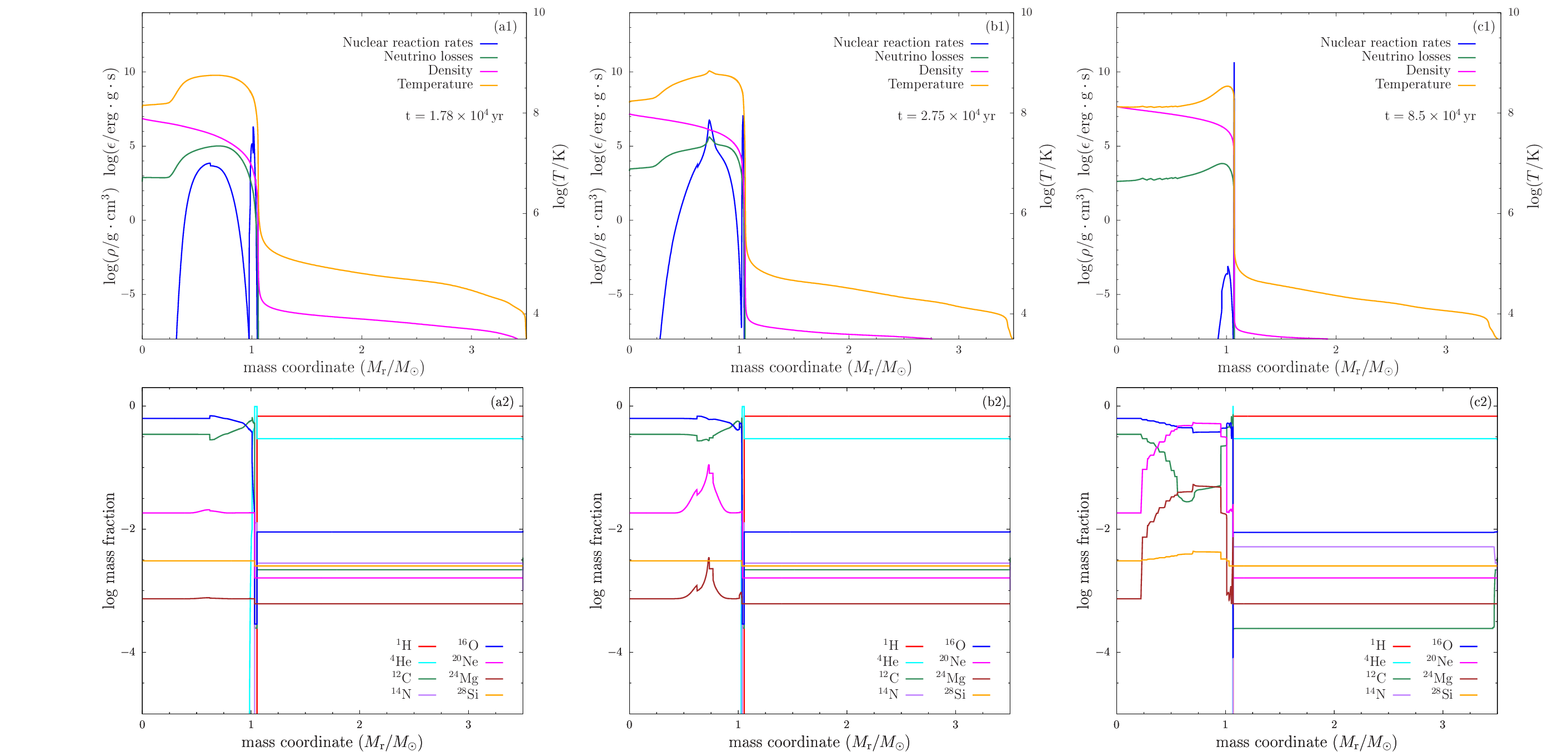,angle=0,width=18.2cm}
 \caption{Similar to Fig.\,2, but for the $3.0{M}_{\odot}$ AGB star + $0.5{M}_{\odot}$ CO WD merger remnant. Panel (a1) and (a2): H/He shell burning phase prior to off-center carbon ignition; Panel (b1) and (b2): at the onset of off-center carbon ignition; Panel (c1) and (c2): carbon shell burning phase.}
  \end{center}
    \label{fig: 4}
\end{figure*}

For the less massive merger remnant, due to the low core mass, the off-center carbon burning is not immediately triggered after the merger. We use the $3.0{M}_{\odot}$ AGB star + $0.5{M}_{\odot}$ CO WD merger remnant as an example, with structure profiles at different evolutionary stages shown in Fig.\,4. At the onset of evolution, the merger core begins to contract, but the mass is insufficient to trigger off-center carbon burning. Instead, the merger remnant begins H/He shell burning. The shell burning transforms $^{\rm 4}{\rm He}$ into $^{\rm 12}{\rm C}$ and $^{\rm 16}{\rm O}$, thereby increasing the mass of the CO core. It takes approximately $2.75\times{10}^{4}$ years of shell burning to increase the CO core mass from $1.0{M}_{\odot}$ to over $1.03{M}_{\odot}$, at which point off-center carbon ignition is triggered at ${M}_{\rm r}\approx0.8{M}_{\odot}$. This core mass and location are consistent with those found in double CO WD merger remnants (e.g., \citealt{2021ApJ...906...53S}). As the carbon flame propagates inward, it does not reach the center but instead quenches at ${M}_{\rm r}\approx0.2{M}_{\odot}$. Subsequently, shell carbon burning begins, marking a transition to multi-shell burning. We stop the calculation when the shell burning commences. We expect that the off-center carbon flame will reignite after the core mass increases further.

\begin{figure*}
\begin{center}
\epsfig{file=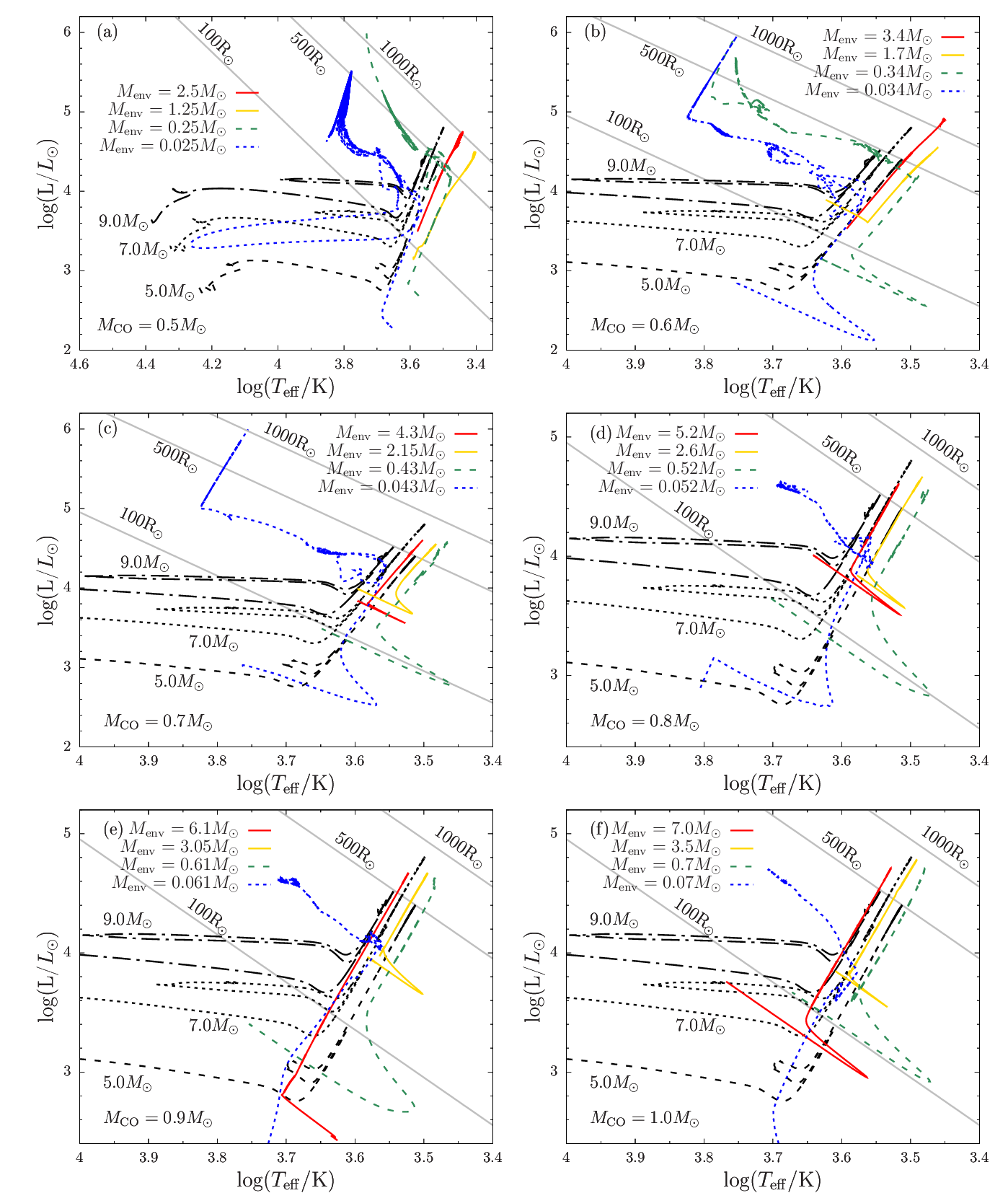,angle=0,width=16.2cm}
 \caption{Evolutionary tracks on the HR diagram for the $24$ merger remnant models. Panel (a)-(f) represent models with different AGB core masses (CO WD masses). In each panel, lines with different colours correspond to models with different envelope masses (while the core mass remains constant). Black dashed, dotted and dash-dotted lines in each panel represent the evolutionary tracks of $5.0{M}_{\odot}$, $7.0{M}_{\odot}$ and $9.0{M}_{\odot}$ stars, respectively.}
  \end{center}
    \label{fig: 5}
\end{figure*}

We computed all $24$ merger models with different core and envelope masses. Their evolution on HR diagram is presented in Fig.\,5. For comparison, we also modeled the evolution of $5.0{M}_{\odot}$, $7.0{M}_{\odot}$ and $9.0{M}_{\odot}$ stars from the MS phase to the AGB phase. For the merger remnants with a similar total mass, their evolution on the HR diagram closely follows that of AGB stars with the same total masses. For merger remnants with the same core mass, the evolutionary tracks on the HR diagram shift toward the lower temperature branch as the envelope mass decreases. This trend similar to the behavior of lower-mass AGB stars. However, as the envelope mass decreases significantly (e.g., if $99\%$ of the envelope mass is lost during the merger process), the evolutionary track on the HR diagram shifts toward the high-mass branch. This is because, as the H-rich envelope becomes extremely thin, the full convective envelope becomes more efficient at transporting energy, resulting in a hotter AGB-like star. Furthermore, as the luminosity increases, the energy transmission efficiency of remnants with extremely thin envelopes increases enough to raise both the temperature and luminosity. The HR diagram of these remnants shows similarities to those of AGB stars, but clear differences arise when comparing remnants with extremely thin envelopes to normal AGB stars.

\begin{figure*}
\begin{center}
\epsfig{file=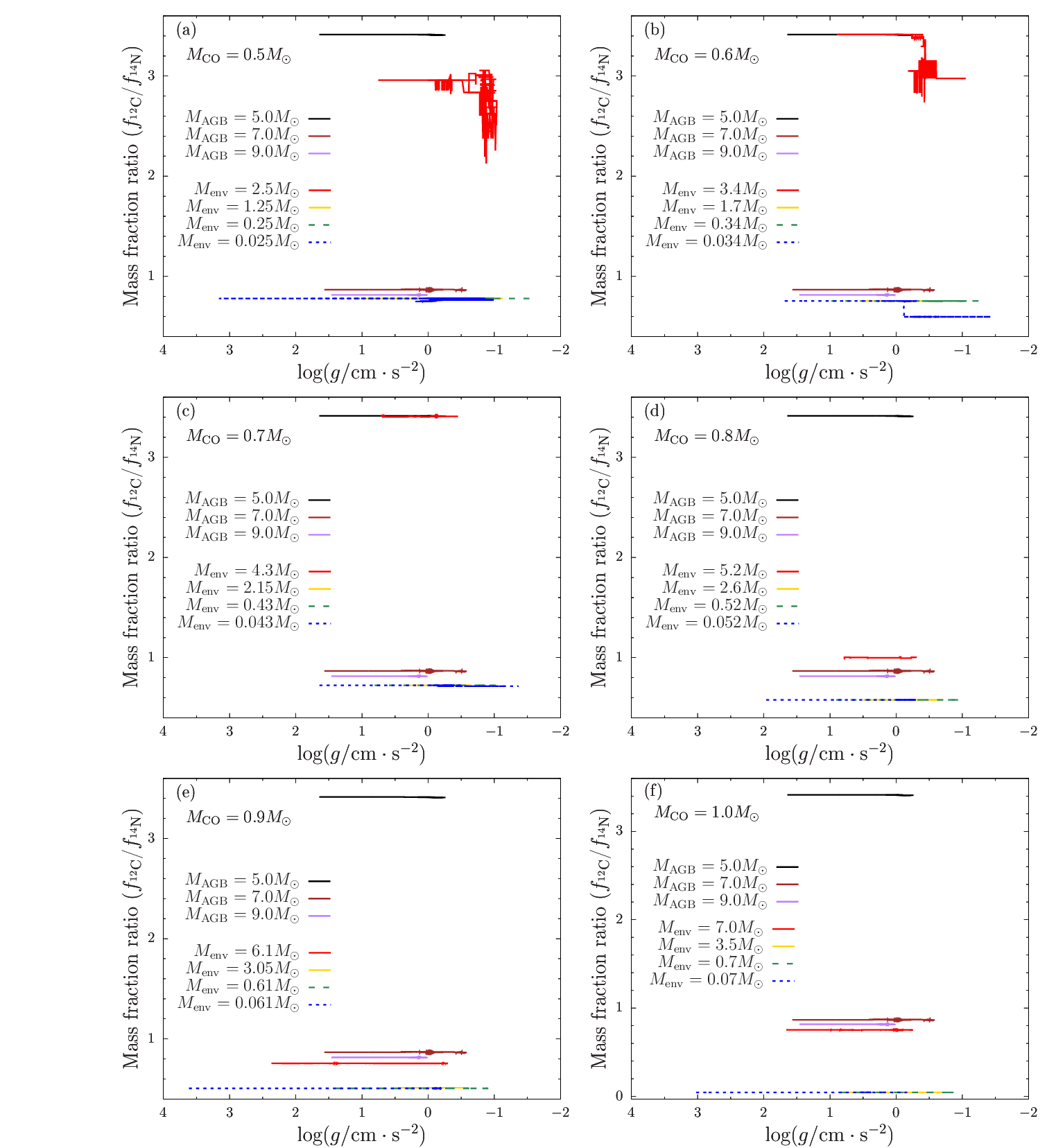,angle=0,width=16.2cm}
 \caption{Similar to Fig.\,5, but showing the evolution of the surface mass fraction ratio of $^{\rm 12}{\rm C}$ to $^{\rm 14}{\rm N}$ as a function of ${\rm log}\,{\rm g}$.}
  \end{center}
    \label{fig: 6}
\end{figure*}

For stars massive than about $5.0{M}_{\odot}$, a blue loop occurs during the He core burning phase. Subsequently, during the AGB phase, hot bottom burning transforms $^{\rm 12}{\rm C}$ into $^{\rm 14}{\rm N}$, potentially reducing the surface $^{\rm 12}{\rm C}$ to $^{\rm 14}{\rm N}$ ratio. This effect may create a discrepancy between the merger remnants and the normal AGB stars. We extracted the information of surface $^{\rm 12}{\rm C}$ to $^{\rm 14}{\rm N}$ mass fraction ratio (${f}_{^{\rm 12}{\rm C}/^{\rm 14}{\rm N}}$) of all $24$ of our merger models and compared them to normal AGB stars. The results are plotted in the ${f}_{^{\rm 12}{\rm C}/^{\rm 14}{\rm N}}$-${\rm log}\,{\rm g}$ diagram (Fig.\,6). For merger remnants with relative low core masses (e.g., core mass lower than $1.4{M}_{\odot}$), the progenitor AGB stars are of relatively low mass, so the merger remnants without mass loss during merger process have nearly the same ${f}_{^{\rm 12}{\rm C}/^{\rm 14}{\rm N}}$ ratio as the corresponding AGB stars. However, remnants with relatively higher core masses originate from more massive AGB progenitors, so they tend to have a lower initial ${f}_{^{\rm 12}{\rm C}/^{\rm 14}{\rm N}}$ ratio. For all models with envelope stripping during the merger process, the remnants show a lower ${f}_{^{\rm 12}{\rm C}/^{\rm 14}{\rm N}}$ ratio compared to their corresponding AGB stars, as hot bottom burning process in the inner regions of the AGB star influences the surface elemental abundance. As a result, the merger remnants that lose part of their H-rich envelope during the merger process typically have a lower ${f}_{^{\rm 12}{\rm C}/^{\rm 14}{\rm N}}$ ratio than AGB stars of similar mass. This conclusion, however, is highly sensitive to the wind mass-loss process during the AGB phase, which we will discuss in more detail in Sect.\,4.

\section{Discussion}

In our fiducial models, we did not account for wind mass-loss process during the post merger evolution of the merger remnants. Mass loss via stellar wind can reduce the mass of the merger remnants over time, potentially altering their evolutionary tracks on the HR diagram and even their final outcomes. More importantly, mass loss could affect their surface elemental abundances, which would subsequently influence their observational properties. To investigate the effects of mass loss, we applied the Blocker wind mass-loss prescription (e.g., \citealt{1995A&A...297..727B} ), given by the formula:
\begin{equation}
    \dot{M}=1.932\times{10}^{-21}{\eta}{{M}^{-3.1}}{{L}^{3.7}}{R}\,({M}_{\odot}/{\rm yr}),
\end{equation}
where, the mass-loss efficiency, ${\eta}=0.05$, is a typical value adopted from the ``${\tt MESA\_default}$''. We applied this mass-loss prescription to both the merger remnants and normal AGB stars. Fig.\,7 shows the evolutionary tracks of the $5.0{M}_{\odot}$ AGB star + $0.7{M}_{\odot}$ CO WD and $7.0{M}_{\odot}$ AGB star + $0.9{M}_{\odot}$ CO WD merger remnants on the HR diagram (Panel (a)), and the evolution of their surface $^{\rm 12}{\rm C}$ to $^{\rm 14}{\rm N}$ ratio (Panel (b)). For comparison, the corresponding evolutionary tracks of $5.0{M}_{\odot}$ and $7.0{M}_{\odot}$ AGB stars are also plotted on Fig.\,7. The wind mass-loss process does not affect the propagation of the off-center carbon flame, i.e., it takes the same amount of time for the flame to reach the center. However, as the envelope is stripped during evolution, the inner H-rich envelope becomes exposed to the surface, where $^{\rm 12}{\rm C}$ is converted into $^{\rm 14}{\rm N}$. This results in a decrease in the surface $^{\rm 12}{\rm C}$ to $^{\rm 14}{\rm N}$ ratio for all evolutionary models, including those of AGB stars. As the evolution progresses, the surface $^{\rm 12}{\rm C}$ to $^{\rm 14}{\rm N}$ ratio tends to converge, meaning that the surface elemental abundance may become similar for both AGB stars and merger remnants. Under the Blocker wind mass-loss prescription, the average mass-loss rate during the evolution of the merger remnants is about $9.0\times{10}^{-6}\,{M}_{\odot}/{\rm yr}$. For remnants with various core and envelope masses, it takes approximately ${10}^{4}$-${10}^{5}$ years for the off-center flame to reach the center, resulting in a loss of about $0.1$-$1.0{M}_{\odot}$ of the envelope during the carbon burning phase. For massive merger remnants, off-center Ne burning occurs shortly after the carbon flame reaches the center, meaning that these remnants still have enough mass to explode as core collapse supernova while remaining on the AGB branch. For less massive remnants, the off-center Ne burning can only be triggered once the core mass increases sufficiently through H/He shell burning. Therefore, the final fate of these remnants depends on the competition between the rate of core mass growth and the mass-loss rate, making their final outcomes uncertain.

\begin{figure*}
\begin{center}
\epsfig{file=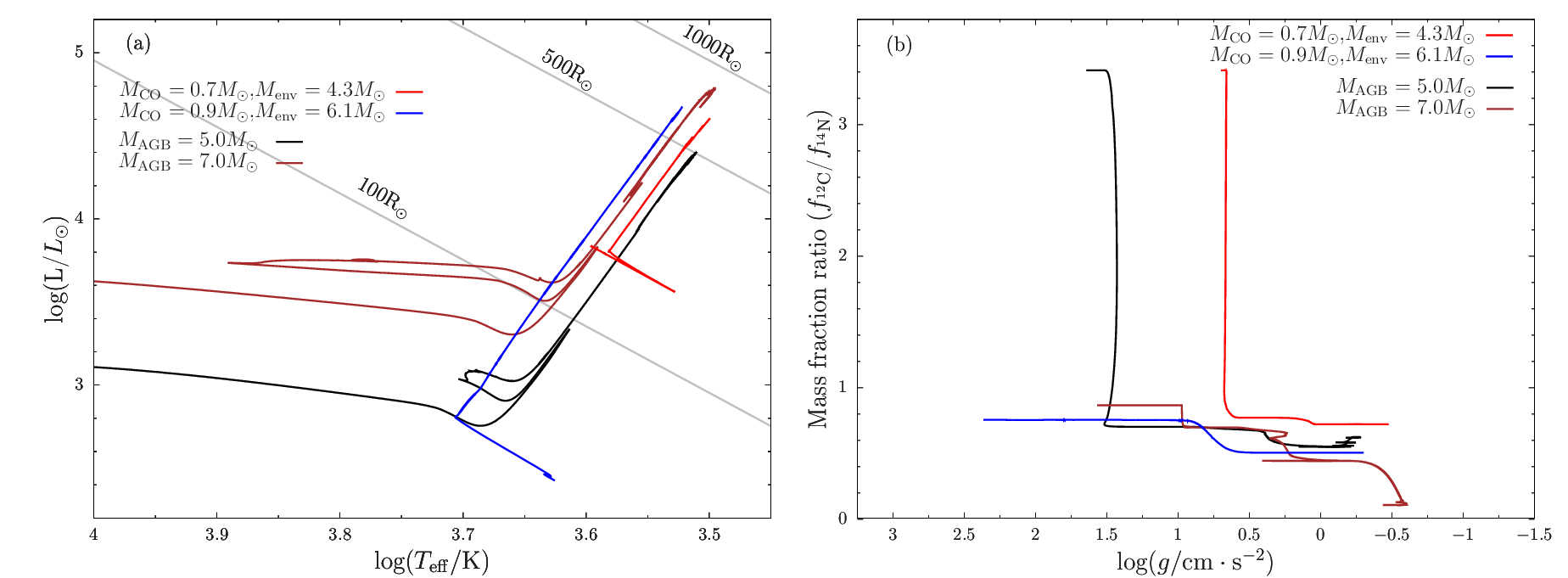,angle=0,width=16.2cm}
 \caption{Evolutionary tracks on the HR diagram (panel a) and the surface mass fraction ratio of $^{\rm 12}{\rm C}$ and $^{\rm 14}{\rm N}$ as a function of ${\rm log}\,{\rm g}$ (panel b) for two sets of models ($5.0{M}_{\odot}$ AGB star + $0.7{M}_{\odot}$ CO WD, red solid line; $7.0{M}_{\odot}$ AGB star + $0.9{M}_{\odot}$ CO WD, blue solid line). For comparsion, the tracks of corresponding single AGB stars with $5.0{M}_{\odot}$ (black solid line) and $7.0{M}_{\odot}$ (brown solid line) are also shown in each panel.}
  \end{center}
    \label{fig: 7}
\end{figure*}

In our fiducial models , we have not considered the impact of rotation during the evolution. The merger of the core of the AGB star with the CO WD may retain some orbital angular momentum, which could cause the merger core to rotate rapidly. The rotation of the core might reduce its internal pressure, potentially delaying the onset of off-center burning. However, previous works have shown that, for double WD merger remnants, the remnant evolves into an H/He-deficient giants after the merger. Even if the angular velocity of the core is initially rapid, the angular momentum of the core quickly transfers to the expanding envelope, causing the rotational velocity of the core to decrease rapidly after the merger (e.g., \citealt{2021ApJ...906...53S}; \citealt{2022MNRAS.512.2972W}; \citealt{2023ApJ...944L..54W}). Thus, as the envelope expands after the merger, the core and envelope quickly reach co-rotation velocity. We therefore estimate that the rotation of the core has a very limited impact on the evolution of the remnant.

Based on our results, if the CE event leads to a merger, the merger remnant shows an increase in luminosity and a decrease in effective temperature, potentially occupying a position on the HR diagram similar to that of SAGB or even RSG stars. Some observed objects, such as HV $2112$, share similar positions to SAGB or RSG stars and are believed to be the result of the merger of degenerate cores with giant stars. HV $2112$ is a bright star in the Small Magellanic Cloud, slightly more luminous than the classical AGB limit, with a relatively long pulsation period (e.g., \citealt{1983ApJ...272...99W}). It is thought to be one of the candidate of Thorne-$\dot{\rm Z}{\rm ytkow}$ objects (${\rm {T\dot{Z}Os}}$), hypothetical unique product of an neutron star merging with an non-degenerate star that forms a single object (e.g., \citealt{1975ApJ...199L..19T}; \citealt{2014MNRAS.443L..94L}; \citealt{2014MNRAS.445L..36T}). These objects may display $irp$-process elements, such as Rb, Mo, etc., although these elements can also be enhanced by the $s$-process in SAGB stars, and Lithium could also be generated through hot bottom burning at the base of the convective envelope. Therefore, it remains difficult to distinguish between a normal giant or a merger remnant based on their surface properties alone. Notably, \cite{2018MNRAS.479.3101B} suggested that HV $2112$ may instead be explained as a $5.0{M}_{\odot}$ AGB star rather than a genuine ${\rm {T}{\dot Z}{O}}$. Based on our results, our most massive merger models pass along the edge of the region occupied by HV $2112$ on the HR diagram (see Fig.\,8), nevertheless an even more massive remnant would be expected to match HV $2112$ more closely. This suggests that some ${\rm {T}{\dot Z}{O}}$ candidates may also be AGB star-WD merger remnants.

\begin{figure*}
\begin{center}
\epsfig{file=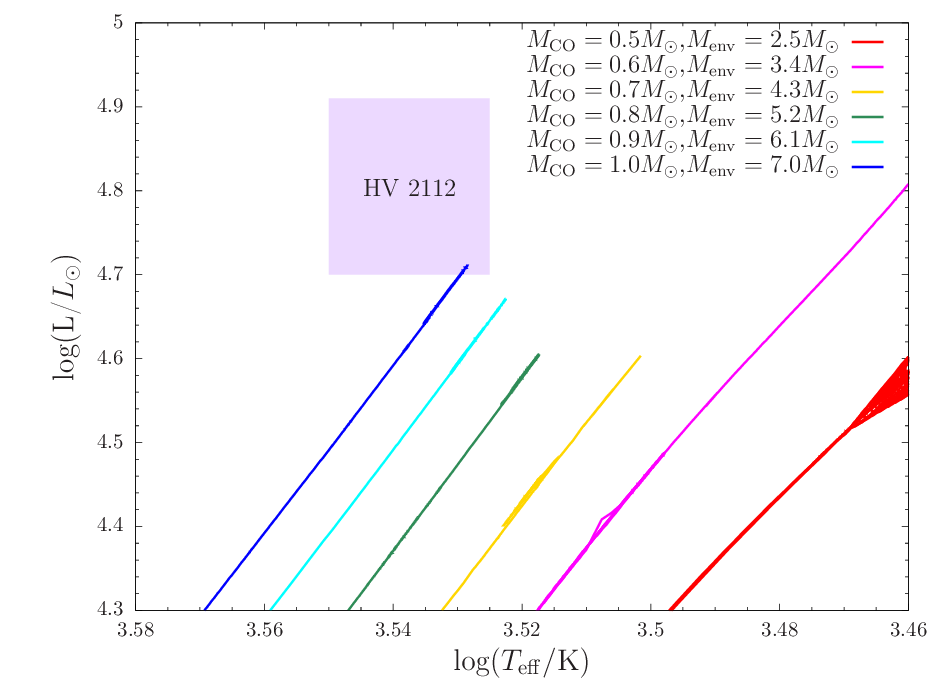,angle=0,width=16.2cm}
 \caption{Evolutionary tracks of our merger models on HR diagram. The purple shaded region marks the observational location of HV $2112$.}
  \end{center}
    \label{fig: 8}
\end{figure*}

Previous works have suggested that asteroseismology could help distinguish the merger remnants from normal giants (e.g., \citealt{2021MNRAS.508.1618R}; \citealt{2022NatAs...6..673L}). These works proposed that merger remnants could be identified through asteroseismic measurements of red giant stars, using gravity mode period spacing together with the asteroseismic mass. However, for the AGB star-WD merger remnants and AGB stars, both typically exhibit large-amplitude, long-period pulsations and strong convective noise in their observed light curves. These components obscure solar-like oscillations, making it difficult to reliably extract asteroseismic parameters such as ${\Delta}{\nu}$ and ${\Delta}{\Pi}$. As a result, unlike RGB or RC stars, AGB stars cannot be effectively diagnosed using mode spacings to constrain their core and envelope masses.

\section{Summary}

In this work, we constructed stellar models for AGB star-CO WD merger remnants, considering different core and envelope masses, and investigated their post merger evolution. We found that the merger remnants undergo off-center carbon burning shortly after the merger. The carbon flame takes approximately ${10}^{4}$-${10}^{5}$ years to reach the center. For massive merger remnants, this may lead to an explosion as a core collapse supernova, while less massive merger remnants will experience core contraction followed by further H shell burning. On the HR diagram, the merger remnants occupy positions similar to normal AGB stars, but those where a significant portion of the H envelope is stripped during the merger process tend to have higher effective temperatures. The surface elemental abundance ratio of $^{\rm 12}{\rm C}$ to $^{\rm 14}{\rm N}$ remains relatively constant during the post merger evolution. When wind mass-loss is considered, the merger remnants and normal AGB stars may have indistinguishable surface $^{\rm 12}{\rm C}$ to $^{\rm 14}{\rm N}$. We propose that some giant-like stars, exhibiting similar observational properties to SAGB stars, such as HV $2112$, a candidate for the ${\rm {T}{\dot Z}{O}}$, might also be merger remnants of an AGB star and WD, suggesting a possible origin, though further investigations are needed to confirm this possibility.

\begin{acknowledgments}

We thank the anonymous referee for very helpful suggestions
on the manuscript. We thank Philipp Podsiadlowski, Yan Li, Meng Sun, Xianfei Zhang, Jie Lin, Zhenwei Li and Zhanwen Han for helpful discussion. This study is supported by the National Natural Science Foundation of China (Nos 12473032, 12288102, 12225304, 12090040/12090043, 12503044, 12433009), the National Key R\&D Program of China (No. 2021YFA1600404), the Yunnan Revitalization Talent Support Program (Young Talent project, Yunling Scholar Project), the Yunnan Fundamental Research Project (Nos 202501AW070001, 202201BC070003, 202301AU070039, 202501AS070005), and the International Centre of Supernovae, Yunnan Key Laboratory (No. 202302AN360001).

\end{acknowledgments}

\bibliography{citation}{}
\bibliographystyle{aasjournal}

\end{document}